\documentstyle[12pt,epsfig]{article}

\def\beq{\begin{equation}}
\def\eeq{\end{equation}}
\def\bea{\begin{eqnarray}}
\def\eea{\end{eqnarray}}
\def\bq{\begin{quote}}
\def\eq{\end{quote}}

\def\bq{\begin{quote}}
\def\eq{\end{quote}}

\begin{document}

\title{\hspace{4.1in}{\small SHEP 01-21; OUTP 0146P\bigskip }\\
\hspace{4.1in}\\
Fermion Masses and Mixing Angles from $SU(3)$ Family Symmetry}
\author{S. F. King$^{\dagger }$and G. G. Ross$^{\ddagger }$ \\
$^{\dagger }$Department of Physics and Astronomy, University of Southampton,%
\\
Southampton, SO17 1BJ, U.K.\\
$^{\ddagger }$Department of Physics, Theoretical Physics, University of
Oxford,\\
1 Keble Road, Oxford OX1 3NP, U.K.}
\date{}
\maketitle

\begin{abstract}
We propose a model based on $SU(3)$ family symmetry which leads to a
successful description of quark and lepton masses and mixing angles
including approximate bi-maximal mixing in the neutrino sector suitable for
the LOW or quasi-vacuum solar solutions, with the atmospheric angle
predicted to be accurately maximal due to the $SU(3)$ symmetry. The model
predicts a CHOOZ angle $\theta _{13}\sim |V_{ub}|.$ The $SU(3)$ symmetry can
also ensure the near degeneracy of squarks and sleptons needed to avoid
large flavour changing neutral currents.
\end{abstract}

\newpage

\section{Introduction}

The origin of fermion masses and mixing angles has been a long standing
puzzle\cite{GGR}. Our determination of the mass matrices has been improved
by new experimental information on both the quark and lepton sectors.
Particularly intriguing is the appearance of an almost maximal neutrino
mixing angle $\theta _{23}=45^{0}$ responsible for atmospheric neutrino
oscillation and the suggestion of a near maximal mixing angle $\theta
_{12}\approx 45^{0}$ in order to describe solar neutrino oscillation\cite
{King:2001uq}. Apart from the CHOOZ angle this contrasts sharply with the
quark mixing angles which are all small.

In this paper we shall show that a near maximal mixing angle may be a signal
of an underlying non-Abelian family symmetry and demonstrate how vacuum
alignment in such a model can lead to maximal mixing. Further we argue that
the quark mass matrix is also indicative of a non-Abelian symmetry, although
in this case the symmetry is responsible for the {\it smallness} of quark
mixing angles. The origin of this apparent contradiction is the see-saw
mechanism which alters the form of the neutrino mass matrix from that of the
quarks and charged leptons.

Non-Abelian family symmetries have been considered before\cite{nonabelian}.
In supersymmetric theories they offer an elegant solution to the flavour
problem, ensuring that the squarks are nearly degenerate and thus
suppressing flavour changing neutral currents (FCNC). Given that the third
family of quarks and leptons is much heavier, and that the dominant
contribution to FCNC comes from the light generations, early attempts to
develop such theories considered the non-Abelian group $SU(2)$ acting on the
first two generations only\cite{bhr}. However these theories do not offer an
easy explanation for the near maximal neutrino mixing in the neutrino sector
involving the second and third generations. Such mixing requires a
correlation between the $(2,2)$ and $(2,3)$ elements of the mass matrix.
This strongly suggests an underlying non-Abelian symmetry involving the
third generation too. For this reason we consider here a $SU(3)$ family
symmetry acting on all three generations with the view to addressing both
these questions. A gauged $SU(3)$ family symmetry which could provide a
dynamical origin for the three observed families of quarks and leptons has a
long history in the literature\cite{su3}. However as far as we are aware
previous $SU(3)$ models have not been able to use the symmetry to relate
mass matrix elements after spontaneous symmetry breaking. Here we show how
this can be done via a novel mechanism for vacuum alignment.

In this paper we shall consider a particular class of $SU(3)$ model which
gives a successful description of quark masses and mixing angles, and
simultaneously gives approximately bi-maximal leptonic mixing with $\theta
_{23}$ fixed to be almost maximal by a combination of $SU(3)$ and vacuum
alignment. In both cases the origin of the mixing angles is controlled by the
non-Abelian vacuum structure relating the second and third generations%
\footnote{%
With an Abelian $U(1)$ family symmetry \cite{abelian} the mixing angle $%
\theta _{23}$ cannot be enforced to be accurately maximal, but it may be
large.}.

It is instructive to see how data favours such a structure in the quark mass
matrix. A recent analysis of all experimental measurements relating to quark
masses and mixings including the latest measurements at BaBar and BELLE has
found the following form for the up and down quark mass matrices \cite
{Roberts:2001zy} 
\begin{equation}
\frac{M^{u}}{m_{t}}=\left( 
\begin{array}{ccc}
0 & b^{\prime }\epsilon ^{3} & c^{\prime }\epsilon ^{3} \\ 
b^{\prime }\epsilon ^{3}e^{i\phi ^{\prime }} & \epsilon ^{2} & a^{\prime
}\epsilon ^{2} \\ 
? & ? & 1
\end{array}
\right)  \label{mu}
\end{equation}

and 
\begin{equation}
\frac{M^{d}}{m_{b}}=\left( 
\begin{array}{ccc}
0 & b\bar{\epsilon}^{3} & c\bar{\epsilon}^{3} \\ 
b\bar{\epsilon}^{3}e^{i\phi } & \bar{\epsilon}^{2} & a\bar{\epsilon}^{2} \\ 
? & ? & 1
\end{array}
\right)  \label{md}
\end{equation}

The equality of the magnitudes of the $(1,2)$ and $(2,1)$ elements leads to
the successful Gatto, Sartori, Tonin (GST) relation\cite{gst} for a $(1,1)$
texture zero. The $(3,1)$ and $(3,2)$ elements are only weakly constrained
because measurement of the quark masses and the CKM matrix does not provide
enough information to determine the full quark mass matrices. The parameters
of the up quark mass matrix $\ $are given by $\epsilon =0.05,$ $b^{\prime
}\simeq 1$ while $a^{\prime }$ and $c^{\prime }$ are very weakly
constrained. The parameters of the down quark mass matrix are much better
determined with 
\begin{eqnarray}
\bar{\epsilon} &=&0.15\pm 0.01,\quad b=1.5\pm 0.1,\quad a=1.31\pm 0.14, 
\nonumber \\
{|c|} &=&{0.4\pm 0.02,\quad \psi =24^{0}\pm 3^{0}} \ \ or \\
{|c|} &=&{1.27\pm 0.05,\quad \psi =-58^{0}\pm 5^{0}}
\end{eqnarray}
where $c=\left| c\right| e^{i\psi }.$ The fact that the $(2,2)$ and $(2,3)$
matrix elements are very similar in magnitude is required by the smallness
of $V_{cb}.$ It suggests a relation between these elements, suggestive of a
non-Abelian symmetry. The recent data now also requires that the $(1,2)$ and 
$(1,3)$ are also quite similar, supporting this interpretation and
disfavouring the promising symmetric texture zero structure with zeroes in
the (1,1) and (1,3) elements \cite{Ramond:1993kv,bhr}\footnote{%
If one allows for an asymmetric form of the quark mass matrices with large
entries below the diagonal it is still possible to have a $(1,3)$ zero\cite
{Roberts:2001zy} . Here we concentrate on symmetric structures which
naturally accommodate the successful GST relation.}.

In the present paper we shall propose a model of fermion masses and mixing
angles based on an $SU(3)$ family symmetry, together with certain discrete
symmetries which are required to forbid unwanted operators. The model we
propose will give rise to Yukawa matrices of the form\footnote{%
Such matrices have been considered before, eg see \cite{altarelli}, but not
with a symmetry ensuring equality of matrix elements.} 
\begin{equation}
Y\approx \left( 
\begin{array}{rll}
O(\epsilon ^{8}) & \lambda \epsilon ^{3} & \lambda \epsilon ^{3} \\ 
-\lambda \epsilon ^{3} & \epsilon ^{2} & \epsilon ^{2} \\ 
-\lambda \epsilon ^{3} & \epsilon ^{2} & 1
\end{array}
\right)  \label{yuk}
\end{equation}
in leading order of the expansion parameter, $\epsilon .$ The coupling $%
\lambda $ is not determined by the symmetry but is expected to be of $O(1).$
The matrix has a hybrid symmetry, being symmetric in the lower block
(3,2)=(2,3) and antisymmetric in the remaining entries (2,1)=-(1,2),
(3,1)=-(1,3). This form applies to the up and down quarks, to the charged
leptons and to he neutrino Dirac couplings between left- and right- handed
neutrino components. In general there is an independent expansion parameter $%
\epsilon $ for each Yukawa matrix. Note that at leading order each Yukawa is
described by only two free parameters! \ The right- handed Majorana coupling
matrix has a different structure.

\section{The $SU(3)$ family symmetry.}

Here we outline the structure of the model, postponing details to later
sections. We start with an $SU(3)$ family symmetry which commutes with the
Standard Model (SM) gauge group. We shall construct a supersymmetric version
of the theory so, apart from the modifications needed to implement the
family symmetry, we have the basic structure of the minimal supersymmetric
SM (MSSM) in which fermions belong to \ chiral supermultiplets. For
simplicity of presentation, we will treat the supersymmetric structure as
implicit. The family symmetry assignments for the left-handed quarks and
leptons are: 
\begin{equation}
\psi _{i}\in (Q_{i},L_{i})\sim {\bf 3},\ \ \psi _{i}^{c}\in
(U_{i}^{c},D_{i}^{c},E_{i}^{c},N_{i}^{c})\sim {\bf 3}  \label{reps}
\end{equation}
where $i=1,2,3$ are $SU(3)$ labels. To build a viable model we need
spontaneous breaking 
\begin{equation}
SU(3)\longrightarrow SU(2)\longrightarrow {\rm Nothing}
\end{equation}

Note that whenever we write $SU(3)$ it will always refer to the new family
symmetry and $SU(2)$ will always refer to its subgroup. These should not be
confused with the SM gauge group factors with which the family symmetry
commutes.

The Higgs fields responsible for the above symmetry breaking are the $SU(3)$
antitriplets (but SM singlets) $\phi _{3}^{i}$, $\phi _{23}^{i}$, which
develop vacuum expectation values (VEVs) 
\begin{equation}
<\phi _{3}>=\left( 
\begin{array}{c}
0 \\ 
0 \\ 
a_{3}
\end{array}
\right) ,\ \ <\phi _{23}>=\left( 
\begin{array}{c}
0 \\ 
b \\ 
b
\end{array}
\right) .
\end{equation}
Note that $<\phi _{3}>$ can always be rotated into the third position using $%
SU(3)$. The alignment of the $<\phi _{23}>$ VEVs is non-trivial and is the
subject of a later section.

The leading contributions to the Yukawa matrices arise from operators which
are quadratic in $\phi _{3}^{i}$ and $\phi _{23}^{i}$ 
\begin{equation}
\left( \frac{1}{M^{2}}\psi _{i}\phi _{23}^{i}\psi _{j}^{c}\phi _{23}^{j}+%
\frac{1}{M_{3}^{2}}\psi _{i}\phi _{3}^{i}\psi _{j}^{c}\phi _{3}^{j}\right)
H_{\alpha }  \label{1}
\end{equation}
where $H_{\alpha }$ are the two Higgs doublets of the MSSM, which are $SU(3)$
singlets, but carry the usual electroweak quantum numbers. If $\psi
^{c}=U^{c},N^{c}$ then $\alpha =2$, while if $\psi ^{c}=D^{c},E^{c}$ then $%
\alpha =1,2.$ The difference between the mass scales $M$ and $M_{3}$ is
discussed below.

\subsection{The operator mass scale\label{expansion}}

The origin of these operators lies in the physics beyond the SM at some
higher energy scale and the inverse mass scale associated with these
operators reflects this higher scale. For example the operator may be due to
a Higgs ``messenger'' sector through mixing of the MSSM Higgs with heavy
vectorlike Higgs fields\cite{froggatt}. In this case the inverse mass scale
is that of the heavy Higgs. Since the heavy Higgs carrying the weak
hypercharge of the $H_{\alpha }$ in the $\alpha =1,2$ sectors may be
different, the case when $\psi ^{c}=U^{c},N^{c}$ is expected to have a
different inverse mass scale, $M=M^{u},$ from the case when $\psi
^{c}=D^{c},E^{c}$ for which $M=M^{d}.$ The operators may also arise through
mixing of the quarks with heavy vectorlike quarks but again we expect the
associated operator mass scale to differ for the up quark and down quark
sectors. When $\phi _{23}$ is replaced by its VEV, the operators of eq(\ref
{1}) generate equal (2,2)=(2,3)=(3,2) elements of $O(b^{2}/M^{2})$ in the 23
block of Eq.\ref{yuk}. As we shall see later this equality is responsible
for the maximal neutrino mixing angle $\theta _{23}$. Due to the different
inverse mass scale the expansion parameters in the up and the down sectors
are expected to be different. We shall denote the expansion parameter in the
up sector by $\epsilon \equiv b/M^{u}$ and that in the down sector by $%
\overline{\epsilon }\equiv b/M^{d}.$

Some comment about the fact that the leading term in Eq.\ref{yuk} for both
the up and the down sectors is of $O(1)$ is in order. At first sight it
would seem that, because the $(3,3)$ term comes from a higher dimension
operator, its magnitude should differ in the up and down sectors for the
same reason that the $O(\epsilon ^{2},\overline{\epsilon }^{2})$ terms
differ. However if $a_{3}>M$, where $M$ is the generic mass scale associated
with the up or the down sector, the expansion parameter is not $a_{3}/M$ but
rather $a_{3}/M_{3}$ where $M_{3}\approx \sqrt{M^{2}+a_{3}^{2}}.$ The reason
for this is that the heavy messenger field responsible for generating the
higher dimension operator necessarily has a contribution to its mass coming
from $a_{3}$ since it couples to $\phi _{3}.$ In addition it will have other
contributions characterised by $M.$ Obviously the largest term dominates
giving the result above. Here we assume that $a_{3}\geq M>b$ so that $\phi
_{3}$ provides the first stage of breaking of $SU(3),$ while the second
stage, triggered by the VEV of $\phi _{2},$ is below the mass scale of the
heavy sector and so generates small effects characterised by the expansion
parameters, $\epsilon $ and $\overline{\epsilon }.$ In what follows we shall
often refer to the expansion parameter as $\epsilon $ but it should be
remembered that this will be $\overline{\epsilon }$ in the down quark
sector. In this paper we shall not consider the details of the heavy sector
responsible for the generation of the operators but content ourselves with
the construction of the effective low energy theory, including operators
consistent with the symmetries of the theory.

\subsection{Subleading operators}

The operators which are mixed in $\phi _{3}^{i}$, $\phi _{23}^{i}$ give the
order $\epsilon ^{3}$ contributions to the (2,3),(3,2),(3,3) elements in Eq.%
\ref{yuk}, and so we require them to be suppressed by $\epsilon ^{2}$
relative to the quadratic operators in Eq.\ref{1}, 
\begin{equation}
\frac{\epsilon ^{2}}{MM_{3}}(\psi _{i}\phi _{23}^{i}\psi _{j}^{c}\phi
_{3}^{j}+\psi _{i}\phi _{3}^{i}\psi _{j}^{c}\phi _{23}^{j})H_{\alpha }
\label{2}
\end{equation}
If the mixed operators were not suppressed then they would imply that the
(2,3) element is of $O(\epsilon ),$ larger than the (2,2) element of $%
O(\epsilon ^{2}),$and would lead to the bad relation $|V_{cb}|\simeq \sqrt{%
m_{s}/m_{b}}$. Here we require a further $O(\epsilon ^{2})$ suppression for
reasons that will become apparent. In order to achieve this suppression we
introduce a discrete symmetry $Z_{2}$ under which $\phi _{3}$ and $\phi
_{23} $ have opposite parity. This allows the quadratic operators, but
forbids the leading order mixed operators. The latter are only allowed at
higher order with the $\epsilon ^{2}$ suppression factor.

In order to generate Yukawa entries in the first row and column, we need to
introduce the $SU(3)$ triplet (SM singlet) Higgs $\overline{\phi }_{3,i}$, $%
\overline{\phi }_{23,i}$, which we shall show develops VEVs along $D-$ and \ 
$F-$flat directions. 
\begin{equation}
<\overline{\phi }_{3}>=\left( 
\begin{array}{c}
0 \\ 
0 \\ 
a_{3}
\end{array}
\right) ,\ \ <\overline{\phi }_{23}>=\left( 
\begin{array}{r}
0 \\ 
b \\ 
-b
\end{array}
\right) .
\end{equation}
The operators responsible for the first row and column of the Yukawa
matrices are then 
\begin{equation}
\frac{\epsilon ^{2}}{M}(\epsilon ^{ijk}\psi _{i}\overline{\phi }_{23,j}\psi
_{k}^{c})H_{\alpha }  \label{3}
\end{equation}
\begin{equation}
\frac{\epsilon ^{6}}{M^{2}M_{3}^{2}}(\epsilon ^{ijk}\psi _{i}\overline{\phi }%
_{3,j}\overline{\phi }_{23,k})(\epsilon ^{lmn}\psi _{l}^{c}\overline{\phi }%
_{3,m}\overline{\phi }_{23,n})H_{\alpha }  \label{4}
\end{equation}
The inclusion of triplet Higgs in addition to the antitriplets introduced
previously allows the antisymmetric tensor $\epsilon ^{ijk}$ to be used to
generate the first row and column of Eq.\ref{yuk}. The operator in Eq.\ref{3}
generates (1,2)=(1,3)=-(2,1)=-(3,1) entries at order $\epsilon ^{3}$, while
the operator in Eq.\ref{4} generates the (1,1) entry at order $\epsilon ^{8}$%
. In the next section we discuss the origin of the required operator
suppression factors.

\section{An $SU(3)$ Model}

\subsection{Operator Analysis}

In order to give a specific realisation of the basic scenario of the
previous section we consider the full (SM singlet) scalar sector of the
model summarised in Table \ref{higgs}. To ensure that the operators are only
allowed at the specified orders we have introduced an R-symmetry under which
all superpotential terms are required to have $R=2$. With these fields and
VEVs introduced in the previous Section, the leading terms generating the
fermion masses allowed by the family symmetries come from the superpotential 
\begin{eqnarray}
P &\sim &\left( \frac{1}{M_{3}^{2}}\psi _{i}\phi _{3}^{i}\psi _{j}^{c}\phi
_{3}^{j}+\frac{1}{M^{2}}\psi _{i}\phi _{23}^{i}\psi _{j}^{c}\phi
_{23}^{j}\right) H_{\alpha }  \label{P41} \\
&+&\frac{1}{M_{3}^{3}M^{3}}(\psi _{i}\phi _{23}^{i}\psi _{j}^{c}\phi
_{3}^{j}+\psi _{i}\phi _{3}^{i}\psi _{j}^{c}\phi _{23}^{j})H_{\alpha }(\phi
_{23}^{k}\overline{\phi }_{23,k})(\phi _{3}^{l}\overline{\phi }_{3,l})
\label{P42} \\
&+&\frac{1}{M_{3}^{2}M^{3}}(\epsilon ^{ijk}\psi _{i}\overline{\phi _{23}}%
_{j}\psi _{k}^{c})H_{\alpha }(\phi _{3}^{l}\overline{\phi }_{23,l})^{2}
\label{P43} \\
&+&\frac{1}{M_{3}^{8}M^{8}}(\epsilon ^{ijk}\psi _{i}\overline{\phi _{3}}_{j}%
\overline{\phi _{23}}_{k})(\epsilon ^{lmn}\psi _{l}^{c}\overline{\phi _{3}}%
_{m}\overline{\phi _{23}}_{n})H_{\alpha }(\phi _{3}^{p}\overline{\phi }%
_{23,p})^{6}  \label{P45} \\
&&+\frac{1}{M_{3}^{5}M^{4}}\left( \epsilon ^{ijk}\psi _{i}^{c}\psi _{j}%
\overline{\phi }_{3,k}(\phi _{3}^{l}\overline{\phi _{23,l}})^{4}+\epsilon
^{ijk}\psi _{i}^{c}\overline{\phi _{3,j}}\overline{\phi _{23,k}}\phi
_{3}^{j}\psi _{j}(\phi _{3}^{l}\overline{\phi _{23,l}})^{3}\right) H_{\alpha
}  \label{Q}
\end{eqnarray}
where the terms in Eqs.\ref{P41}-\ref{P45} correspond to the operators in
Eqs.\ref{1},\ref{2},\ref{3},\ref{4}. They generate the correct suppression
factors once the fields are replaced by their VEVs appropriately. Eq.(\ref{Q}%
) includes a further subleading term that will be important in the following
discussion.

\begin{table}[tbp] \centering%
%
\begin{tabular}{|ccccc|}
\hline
${\bf Field}$ & ${\bf SU(3)}$ & ${\bf Z}_{2}$ & ${\bf R}$ & ${\bf L}$ \\ 
\hline
$\phi _{3}$ & $\overline{3}$ & $+$ & $+1$ & $0$ \\ 
$\overline{\phi _{3}}$ & ${3}$ & $+$ & $-2$ & $0$ \\ 
$\overline{\phi _{2}}$ & ${3}$ & $+$ & $0$ & $0$ \\ 
$\phi _{23}$ & $\overline{3}$ & $-$ & $+1$ & $0$ \\ 
$\overline{\phi _{23}}$ & ${3}$ & $+$ & $0$ & $0$ \\ 
$\nu $ & $\overline{3}$ & $+$ & $+1$ & $1$ \\ 
$\overline{\nu }$ & ${3}$ & $+$ & $+1$ & -$1$ \\ 
$U$ & $1$ & $-$ & $1$ & $0$ \\ 
$X$ & $1$ & $+$ & $+1$ & $0$ \\ 
$Y$ & $1$ & $+$ & $+2$ & $0$ \\ 
$Z$ & $1$ & $-$ & $0$ & $1$ \\ \hline
\end{tabular}
\caption{Transformation of the Higgs superfields under the family
symmetries.\label{higgs}}%
\end{table}%
%

It is straightforward to show that the operators in $P$ are the leading ones
allowed by the symmetry. In Table \ref{forbidden} we list the operators of
leading dimension which would spoil the form of eq(\ref{yuk}), together with
the symmetry that forbids it. As we shall see in the next section, we may
take the $\overline{\phi }_{2}$VEV to be very small and so operators
involving this field may be ignored. The remaining operators are forbidden
by the family symmetries.

\begin{table}[tbp] \centering%
%
\hfil
\begin{tabular}{cll}
\hline\hline
Category & Forbidden Operator & Reason \\ \hline\hline
&  &  \\ 
& $\psi _{i}\phi _{23}^{i}\psi _{j}^{c}\phi _{3}^{j}H_{\alpha }$ & $Z_{2}$
\\ 
I & $\psi _{i}\phi _{23}^{i}\psi _{j}^{c}\phi _{3}^{j}H_{\alpha }(\bar{\phi}%
_{3l}\phi _{3}^{l})$ & $Z_{2}$,$R$ \\ 
& $\psi _{i}\phi _{23}^{i}\psi _{j}^{c}\phi _{3}^{j}H_{\alpha }(\bar{\phi}_{{%
23}l}\phi _{3}^{l})$ & $Z_{2}$,$R$ \\ 
& $\psi _{i}\phi _{23}^{i}\psi _{j}^{c}\phi _{3}^{j}H_{\alpha }(\bar{\phi}%
_{3l}\phi _{23}^{l})$ & $R$ \\ 
&  &  \\ \hline
&  &  \\ 
& $\epsilon ^{ijk}\psi _{i}\bar{\phi}_{{23}j}\psi _{k}^{c}H_{\alpha }$ & $R$
\\ 
II & $\epsilon ^{ijk}\psi _{i}\bar{\phi}_{{23}j}\psi _{k}^{c}H_{\alpha }(%
\bar{\phi}_{3l}\phi _{3}^{l})$ & $R$ \\ 
& $\epsilon ^{ijk}\psi _{i}\bar{\phi}_{{23}j}\psi _{k}^{c}H_{\alpha }(\bar{%
\phi}_{{23}l}\phi _{3}^{l})$ & $R$ \\ 
& $\epsilon ^{ijk}\psi _{i}\bar{\phi}_{{23}j}\psi _{k}^{c}H_{\alpha }(\bar{%
\phi}_{3l}\phi _{23}^{l})$ & $Z_{2}$,$R$ \\ 
&  &  \\ \hline
&  &  \\ 
& $\epsilon ^{ijk}\psi _{i}\bar{\phi}_{3j}\psi _{k}^{c}H_{\alpha }$ & $R$ \\ 
III & $\epsilon ^{ijk}\psi _{i}\bar{\phi}_{3j}\psi _{k}^{c}H_{\alpha }(\bar{%
\phi}_{3l}\phi _{3}^{l})$ & $R$ \\ 
& $\epsilon ^{ijk}\psi _{i}\bar{\phi}_{3j}\psi _{k}^{c}H_{\alpha }(\bar{\phi}%
_{{23}l}\phi _{3}^{l})$ & $R$ \\ 
& $\epsilon ^{ijk}\psi _{i}\bar{\phi}_{3j}\psi _{k}^{c}H_{\alpha }(\bar{\phi}%
_{3l}\phi _{23}^{l})$ & $Z_{2}$,$R$ \\ 
&  &  \\ \hline
&  &  \\ 
IV & $(\epsilon ^{ijk}\psi _{i}\bar{\phi}_{{23}j}\bar{\phi}_{3k})(\psi
_{l}^{c}\phi _{3}^{l})H_{\alpha }$ & $R$ \\ 
& $(\epsilon ^{ijk}\psi _{i}\bar{\phi}_{{23}j}\bar{\phi}_{3k})(\psi
_{l}^{c}\phi _{23}^{l})(\bar{\phi}_{3m}\phi _{23}^{m})H_{\alpha }$ & $R$ \\ 
& $(\epsilon ^{ijk}\psi _{i}\bar{\phi}_{{23}j}\bar{\phi}_{3k})(\psi
_{l}^{c}\phi _{23}^{l})H_{\alpha }$ & $Z_{2}$,$R$ \\ 
&  &  \\ \hline
&  &  \\ 
V & $(\epsilon ^{ijk}\psi _{i}\bar{\phi}_{{23}j}\bar{\phi}_{3k})(\epsilon
^{lmn}\psi _{l}^{c}\bar{\phi}_{{23}m}\bar{\phi}_{3n})H_{\alpha }$ & $R$ \\ 
&  &  \\ \hline\hline
\end{tabular}
\caption{\footnotesize{Forbidden operators and the reason why they are 
excluded.}\label{forbidden}}%
\end{table}%
%

\subsection{Vacuum Alignment\label{vac}}

The critical feature of the model is the vacuum alignment which arranges for
the fields $\phi _{23},$ $\overline{\phi }_{23}$ to acquire VEVs of equal
magnitude in the $2$ and $3$ directions. Here we discuss how this comes
about. The initial stage of symmetry breaking is triggered by VEVs for $\phi
_{3}$ and $\overline{\phi _{2}}.$ We assume that the fields have Yukawa
couplings to the heavy sector fields which drive soft mass squared terms
negative at some scale, $\Lambda ,$ through radiative corrections. The VEV
cannot be larger than this scale as the effective potential has the form $%
m^{2}(\phi ^{2})\phi ^{2}$ and clearly is positive above this scale and
negative below. We suppose that such radiative effects trigger VEVs for $%
\phi _{3}$ and $\overline{\phi _{2}}.$ Without loss of generality we can
choose the basis such that $\phi _{3}^{T}=\left( 
\begin{array}{ccc}
0 & 0 & a_{3}
\end{array}
\right) .$ The alignment of the VEVs of $\phi _{3}$ and $\overline{\phi _{2}}
$ is due to the term in the superpotential 
\begin{equation}
P_{1}\sim X\phi _{3}\overline{\phi _{2}}  \label{P1}
\end{equation}
such that along the F-flat direction $\left| F_{X}\right| =0,$ $\phi _{3}$
and $\overline{\phi _{2}}$ are orthogonal. Again without loss of generality
we can choose the basis such that $\overline{\phi _{2}}=\left( 
\begin{array}{ccc}
0 & a_{2} & 0
\end{array}
\right) .$ As we demonstrate in the Appendix, for a particular range of
parameters, $D-$ flatness requires $\overline{\phi _{3}}=\left( 
\begin{array}{ccc}
0 & 0 & a_{3}
\end{array}
\right) .$

Consider now the field $\phi _{23}.$ We assume that due to different heavy
sector interactions, its mass squared remains positive and its VEV is
triggered by an F-term. The $SU(3)$ symmetry guarantees the equality of the
components of the soft mass term $m_{23}^{2}\left| \phi _{23}\right|
^{2}=m_{23}^{2}(\left| \phi _{23,1}\right| ^{2}+\left| \phi _{23,2}\right|
^{2}+\left| \phi _{23,3}\right| ^{2}).$ We will use this to obtain the
vacuum alignment required. Consider the superpotential terms consistent
with the symmetries of Table \ref{higgs}. 
\begin{equation}
P_{2}\sim Y(\phi _{23}\overline{\phi _{2}}\phi _{23}\overline{\phi _{3}}-\mu
^{4}), \ \ 
P_{3}\sim
U\phi _{23}\overline{\phi _{23}}  \label{P2}
\end{equation}
Here $\mu $ is a mass scale associated with spontaneous symmetry breaking in
the heavy sector. Requiring $\left| F_{Y}\right| =0$ implies $<\phi
_{23,2}\phi _{23,3}>=\mu ^{4}/(a_{2}a_{3})\equiv b^{2}$ for $m_{23}^{2}>0%
\footnote{%
We take $b$ to be real, although our analysis is not sensitive to this
assumption.}.$ This is the vacuum alignment that will lead to maximal mixing
in the neutrino sector. The resultant pattern of symmetry breaking along $%
m_{23}^{2}$ $D-$ and $F-$flat directions, for the range of parameters
detailed in the Appendix, is given in Table $\ref{vevs}$. Note that the fact
the magnitude of the two VEVs in $\phi _{23},$ $\overline{\phi }_{23}$ are
equal follows from the minimisation of $m_{23}^{2}\left| \phi _{23}\right|
^{2}$ and that it is the underlying $SU(3)$ which ensures this by ensuring
the soft mass terms are equal for all three components.

Finally we consider the fields 
\footnote{These lepton number violating 
Higgs superfields should not be confused 
with neutrino superfields.} 
$\nu ,$ $\overline{\nu }$ responsible for the
Majorana masses.
Below the scale of $SU(3)$ breaking, $a_{3}$, only the $%
SU(2)$ gauge bosons are light and contribute to the running of the $%
\widetilde{\nu }$ masses. Thus, below this scale, $\widetilde{\nu }_{1,2}$
become heavier than $\widetilde{\nu }_{3}.$ However $\nu $ may also have
Yukawa couplings to heavy states, which need not feel large $SU(3)$ breaking
if they do not couple to $\phi _{3}$ in the superpotential. Such terms will
drive $m_{\nu }^{2}$ negative in the usual way, driving a VEV in the
lightest $\widetilde{\nu }_{3}$. Thus it is easy to align the VEV of $\nu $
without the addition of any superpotential ``alignment'' terms. However, if
such terms are present they will skew the VEV from the direction favoured by
the soft terms because such an F-term is not suppressed by the SUSY breaking
scale that characterises the soft mass terms. When we come to consider
neutrino masses it will be necessary to generate a component of the VEV
along the $\nu _{1}$ direction. It is easy to construct just such a term via
the introduction of an $F-$term consistent with the symmetries. Suppose we
have the field $Z$ with quantum numbers shown in Table $\ref{vevs}$. The
allowed superpotential terms are 
\begin{equation}
P_{4}\sim
Z\nu ^{i}(\epsilon _{ijk}\phi _{23}^{j}\phi _{3}^{k}(\phi _{3}\overline{\phi 
}_{3})+\overline{\phi }_{23i}(\phi _{3}\overline{\phi }_{23})^{2}(\overline{%
\phi }_{3}\phi _{23}))
\end{equation}
Now the vanishing of $F_{Z}$ implies <$\nu ^{1}-\epsilon _{\nu
}^{2}\nu ^{3}>=0$ so the VEV is skewed in the direction <$\nu
>=(\epsilon _{\nu }^{2},0,1)\sigma .$

\begin{table}[tbp] \centering%
%
\begin{tabular}{|cc|}
\hline
${\bf Field}$ & ${\bf VEV}$ \\ 
$\phi _{3}^{T}$ & $\left( 
\begin{array}{ccc}
0 & 0 & a_{3}
\end{array}
\right) $ \\ 
$\overline{\phi _{3}}$ & $\left( 
\begin{array}{ccc}
0 & 0 & a_{3}
\end{array}
\right) $ \\ 
$\overline{\phi _{2}}$ & $\left( 
\begin{array}{ccc}
0 & a_{2} & 0
\end{array}
\right) $ \\ 
$\phi _{23}^{T}$ & $\left( 
\begin{array}{ccc}
0 & b & b
\end{array}
\right) $ \\ 
$\overline{\phi _{23}}$ & $\left( 
\begin{array}{ccc}
0 & b & -b
\end{array}
\right) $ \\ 
$\nu ^{T}$ & $\left( 
\begin{array}{ccc}
0 & 0 & \sigma
\end{array}
\right) $ \\ 
$\overline{\nu }$ & $\left( 
\begin{array}{ccc}
0 & 0 & \sigma
\end{array}
\right) $ \\ \hline
\end{tabular}
\caption{Vacuum expectation values. Phases are not shown\label{vevs}}%
\end{table}%
%

\subsection{Quark masses}

The operators of eqs(\ref{P41}-\ref{P45}) with the VEVs of Table \ref{vevs}
generate the Yukawa matrices of the form given in eq(\ref{yuk}). This is of
the correct form to reproduce the ``experimental'' form for the mass
matrices given in eqs(\ref{mu}) and (\ref{md}), once the appropriate
expansion parameters are inserted as discssed in Section \ref{expansion}.
Note that the leading operators require$\left| b^{\prime }\right| =\left|
c^{\prime }\right| ,\left| b\right| =\left| c\right| ,$ $\left| a^{\prime
}\right| =1,$ $|a|=1.$ This follows from the vacuum alignment of the $\phi
_{23},$ $\overline{\phi }_{23}$ fields. However subleading operators will
spoil these relations. Keeping all the contributions in eqs(\ref{P41}-\ref{Q}%
) we obtain the form 
\begin{equation}
\frac{M}{m_{3}}=\left( 
\begin{array}{ccc}
\epsilon ^{8} & \lambda \epsilon ^{3}+O(\epsilon ^{4}) & \lambda \epsilon
^{3}+O(\epsilon ^{4}) \\ 
-\lambda \epsilon ^{3}+O(\epsilon ^{4}) & \epsilon ^{2} & \epsilon
^{2}+\lambda ^{\prime }\epsilon ^{3} \\ 
-\lambda \epsilon ^{3}+O(\epsilon ^{4}) & \epsilon ^{2}+\lambda ^{\prime
}\epsilon ^{3} & 1+\epsilon ^{2}+2\lambda ^{\prime }\epsilon ^{3}
\end{array}
\right)  \label{nonabelianmatrix}
\end{equation}
For the case of the down quark mass matrix the expansion parameter is not
very small, $\overline{\epsilon }\simeq 0.15$ and so the corrections can be
quite large. Using this freedom we see that the $SU(3)$ model is able to
reproduce eqs(\ref{mu}) and (\ref{md}) and thus the quark masses and mixing
angles.

\subsection{Charged lepton masses}

With the lepton assignments to $SU(3)$ triplets as in eq(\ref{reps}), and
assuming the same $Z_{2}$ and $R$ charges as for the quarks, the lepton mass
matrix will have the same form as the of the quarks, although the expansion
parameter may differ if the mass scale associated with the operators is
different. However, since the charged leptons get their mass from the same
Higgs as generates the down quark masses, if the operators are dominantly
generated by mixing to Heavy Higgs fields the expansion parameter for down
quarks and charged leptons will be the same. The same will be true if the
heavy charged leptons have the same mass scale as the heavy down quarks.
Here we assume this is the case and so one should use the expansion
parameter $\overline{\epsilon }$ in eq(\ref{nonabelianmatrix}) when
computing the lepton masses. However it is not consistent to have identical
down and charged lepton mass matrices. The relation $m_{b}=m_{\tau }$ is
consistent with the measured values after radiative corrections are
included, provided this equality applies at a very high scale (the GUT
scale?). However to describe the lighter generations it is necessary to have
the approximate relations $m_{s}\simeq m_{\mu }/3$ and $m_{d}\simeq 3m_{e}$
at the high scale. This can be achieved through the choice of different
Yukawa couplings for the charged leptons and down quarks and it may be that
this choice follows from an underlying GUT as suggested by Georgi and
Jarlskog \cite{GJ}. Since we are not attempting here to construct the
underlying theory valid at a high (GUT)\ scale, we will not pursue this
possibility further but merely note that the choice of lepton Yukawa matrix
of the form 
\begin{equation}
Y_{l}=\left( 
\begin{array}{ccc}
\overline{\epsilon }^{8} & \lambda \overline{\epsilon }^{3} & \lambda 
\overline{\epsilon }^{3} \\ 
-\lambda \overline{\epsilon }^{3} & 3\overline{\epsilon }^{2} & 3\overline{%
\epsilon }^{2} \\ 
-\lambda \overline{\epsilon }^{3} & 3\overline{\epsilon }^{2} & 1
\end{array}
\right)  \label{Ye}
\end{equation}
provides an excellent fit to the lepton masses. The form is that required by
the underlying $SU(3)$ symmetry and we have included a factor $3$ to achieve
the desired relations for the two light generations while keeping the other
couplings the same as for the down quarks.

\subsection{Neutrino masses}

We turn now to neutrino masses. As we shall discuss the choice of vacuum
alignment given above allows for near bi-maximal mixing even with the
universal form for the Dirac masses! Due to the possibility that neutrinos
have Majorana masses, the form of the light neutrino mass matrix is expected
to differ substantially from that of the quarks and charged leptons. The
general form of the effective mass matrix is given by the see-saw form 
\begin{equation}
m_{eff}=m_{LR}.M_{RR}^{-1}.m_{LR}^{T}  \label{seesaw}
\end{equation}
where $M_{RR}$ is the $3\times 3$ matrix of Majorana masses for the three
generations of right-handed neutrinos and $m_{LR}=Y_{\nu }v_{2},$ where $%
v_{2}$ is the VEV of the second Higgs doublet, and $Y_{\nu }$ is the
neutrino Yukawa matrix coupling the left- to right- handed neutrino
components. Here we assume there are no $I_{W}=1$ Higgs fields giving rise
directly to a Majorana mass for the left-handed neutrinos.

\subsubsection{The Dirac mass matrix}

Given the neutrino $SU(3)$ assignments of eq(\ref{reps}) we see that the
Dirac mass matrix must be of the same form, eq(\ref{nonabelianmatrix}) as
the charged lepton and quarks. However the expansion parameter may of course
be different. Since in this case the neutrinos get their Dirac mass from the
same Higgs as generates the up quark masses, if the operators are dominantly
generated by mixing to Heavy Higgs fields the expansion parameter for up
quarks and neutrinos will be the same. The same will be true if the heavy
neutrinos have the same mass scale as the heavy up quarks. We assume this is
the case here. As a result the Yukawa matrix is given by 
\begin{equation}
Y_{\nu }=\left( 
\begin{array}{ccc}
\epsilon ^{8} & \lambda _{\nu }\epsilon ^{3}+\lambda _{\nu }^{\prime \prime
}\epsilon ^{4} & \lambda _{\nu }\epsilon ^{3}+\lambda _{\nu }^{\prime \prime
\prime }\epsilon ^{4} \\ 
-\lambda _{\nu }\epsilon ^{3}+\lambda _{\nu }^{\prime \prime }\epsilon ^{4}
& a_{\nu }\epsilon ^{2} & a_{\nu }\epsilon ^{2}+\lambda _{\nu }^{\prime
}\epsilon ^{3} \\ 
-\lambda _{\nu }\epsilon ^{3}+\lambda _{\nu }^{\prime \prime \prime
}\epsilon ^{4} & a_{\nu }\epsilon ^{2}+\lambda _{\nu }^{\prime }\epsilon ^{3}
& 1
\end{array}
\right)  \label{Ynu}
\end{equation}
where we have included the couplings of $O(1).$

\subsubsection{The Majorana mass matrix}

The dominant (heavy right-handed) Majorana mass term comes from 
\begin{equation}
P_{5}\sim \frac{1}{M}\psi _{i}^{c}\nu ^{i}\nu ^{j}\psi _{j}^{c}
\end{equation}
which gives Majorana mass, $M_{3}=\sigma ^{2}/M,$ to the third family and
mass $M_{1}=\epsilon _{\nu }^{4}\sigma ^{2}/M,$ to the first family. There
are further Majorana masses generated higher dimension operators allowed by
the symmetries. The leading ones are 
\begin{equation}
P_{6}\sim \frac{1}{M_{3}M^{5}}\psi _{i}^{c}\phi _{23}^{i}\phi _{23}^{j}\psi
_{j}^{c}\nu ^{k}\overline{\phi _{3,k}}\nu ^{l}\overline{\phi _{23,l}}
\end{equation}
leading to the final form for the (heavy right-handed) Majorana mass matrix
given by 
\begin{equation}
\frac{M_{RR}}{M_{RR,33}}=\left( 
\begin{array}{ccc}
\epsilon _{\nu }^{4} & 0 & \epsilon _{\nu }^{2} \\ 
0 & \epsilon _{\nu }^{3} & \epsilon _{\nu }^{3} \\ 
\epsilon _{\nu }^{2} & \epsilon _{\nu }^{3} & 1
\end{array}
\right)
\end{equation}
The result of this is to give (heavy right-handed) Majorana masses, $%
M_{1}:M_{2}:M_{3}=$ $\epsilon _{\nu }^{4}:\epsilon _{\nu }^{3}:1\ $where we
have allowed for a new expansion parameter $\epsilon _{\nu }.$ Note that if $%
\epsilon _{\nu }<<\epsilon $ the mixing in the light neutrino sector will be
dominated by the Dirac mass matrix. In this case the three Majorana mass
eigenstates are along the $1,$ $2$ and $3$ directions in the basis in which $%
M_{D}$ is written. This alignment proves crucial in allowing for near
bi-maximal mixing.

\subsubsection{The effective mass matrix for the light neutrinos}

The see-saw mechanism gives a light effective Majorana matrix via the
see-saw formula of eq(\ref{seesaw}). The resulting Lagrangian giving the
light doublet neutrino masses is then 
\begin{eqnarray}
{\cal L} &\approx &\frac{(\epsilon ^{8}\nu _{e}+\left( -\lambda _{\nu
}\epsilon ^{3}+\lambda _{\nu }^{\prime \prime }\epsilon ^{4}\right) \nu
_{\mu }+\left( -\lambda _{\nu }\epsilon ^{3}+\lambda _{\nu }^{\prime \prime
\prime }\epsilon ^{4}\right) \nu _{\tau })^{2}}{M_{1}}v^{2}  \nonumber \\
&&+\frac{(\left( a_{\nu }\epsilon ^{2}+\lambda _{\nu }^{\prime }\epsilon
^{3}\right) \nu _{\tau }+a_{\nu }\epsilon ^{2}\nu _{\mu }+\lambda _{\nu
}\epsilon ^{3}\nu _{e})^{2}}{M_{2}}v^{2}  \nonumber \\
&&+\frac{(\nu _{\tau }+\left( a_{\nu }\epsilon ^{2}+\lambda _{\nu }^{\prime
}\epsilon ^{3}\right) \nu _{\mu }+\lambda _{\nu }\epsilon ^{3}\nu _{e})^{2}}{%
M_{3}}v^{2}  \label{neutrino mass}
\end{eqnarray}
Maximal atmospheric neutrino mixing follows if we have single right-handed
neutrino dominance \cite{SRHND} by choosing the parameter $\epsilon _{\nu }$
such that the first right-handed neutrino gives the dominant contribution to
the 23 block of the light effective Majorana matrix. The condition for this
is $\epsilon _{\nu }<\epsilon ^{2}<1$ and is sufficient to guarantee that
the contributions from the second term along the $2$ direction in eq(\ref
{neutrino mass}) is the next most important and determines the mass and
mixing of the second heaviest neutrino state. The resulting mass matrix has
heaviest state $\nu _{a}=(\nu _{\mu }+\nu _{\tau })/\sqrt{2}$ with mass $%
\propto \epsilon ^{6}/\epsilon _{\nu }^{4}.$ Up to the correction terms of
order $\epsilon $ this is {\it maximally} mixed (45$^{0})$ due to the vacuum
alignment of $\phi _{23}^{T}.$

Turning to the lighter neutrinos the second term in eq(\ref{neutrino mass})
provides the dominant contribution if $\epsilon _{\nu }<\epsilon ^{2}.$ A
novel feature of the $SU(3)$ structure is that it leads to the possibility
of near maximal mixing in solar neutrino oscillation too! Due to the $SU(3)$
symmetry the leading contribution from this term is $a_{\nu }^{2}\epsilon
^{4}(\nu _{\mu }+\nu _{\tau })^{2}/M_{2}$ and thus just adds to the $\nu
_{a} $ mass. At subleading order it generates a mass $m_{2}\propto \epsilon
^{6}/\epsilon _{\nu }^{3}$ to an orthogonal component to $\nu _{a}$ which is
a mixture of $\nu _{\mu }-\nu _{\tau }$ and $\nu _{e}$ at the {\it same}
order in $\epsilon .$ As a result there will be large mixing in this sector
too. The ratio of neutrino masses is given by 
\begin{equation}
\frac{m_{2}}{m_{3}}\sim \epsilon _{\nu }<\epsilon ^{2}  \label{ratio}
\end{equation}

With this ratio our model may be consistent with either the LOW or
quasi-vacuum oscillation solutions but not with the LMA solution. With only
two right-handed neutrinos effectively contributing, with the first one
being dominant, and the second being subdominant, we may use the analytic
results of \cite{SRHND2} to estimate the remaining mixing angles in this
model, which are all expressed in terms of neutrino Yukawa matrix elements
in Eq.\ref{Ynu} . We have already seem that the atmospheric mixing angle $%
\theta _{13}$ is given by

\begin{equation}
\tan \theta _{23}=\frac{Y_{21}^{\nu }}{Y_{31}^{\nu }}=1  \label{atmos}
\end{equation}
with small corrections of order $|V_{cb}|$ from the charged lepton sector.
The contribution to the CHOOZ angle from the neutrino sector $\theta
_{13}^{\nu }$ is predicted to be vanishingly small 
\begin{equation}
\theta _{13}^{\nu }=\frac{Y_{11}^{\nu }}{\sqrt{{Y_{21}^{\nu }}^{2}+{%
Y_{31}^{\nu }}^{2}}}\sim \epsilon ^{5}
\end{equation}
Therefore the CHOOZ angle will originate from the charged lepton sector and
from eq.(\ref{Ye}) we predict a CHOOZ angle 
\begin{equation}
\theta _{13}\sim |V_{ub}|\sim ({\overline{\epsilon }})^{3}
\sim 3\times 10^{-3}.
\label{cho}
\end{equation}

The solar angle $\theta _{12}$ is predicted to originate mainly from the
neutrino sector, with small corrections of order the Cabibbo angle from the
charged lepton sector. Neglecting the small CHOOZ angle, and inserting the
maximal atmospheric angle, we find 
\begin{equation}
\tan \theta _{12}=\frac{\sqrt{2}Y_{12}^{\nu }}{Y_{22}^{\nu }-Y_{23}^{\nu }}%
\sim O(1)  \label{solar}
\end{equation}
where the leading contribution to ${Y_{22}^{\nu }-Y_{23}^{\nu }}$ cancels
due to vacuum alignment, and the subleading $\epsilon ^{3}$ term is of the
same order as $Y_{12}^{\nu }$ but comes from a different operator with an
independent coefficient, so the solar angle is not precisely predicted but
is expected to be large.

Note that the large solar mixing angle is due to the equality, in leading
order, of the $(2,2)$ and $(2,3)$ matrix elements of $Y^{\nu }$, which
followed from the vacuum alignment made possible by the underlying $SU(3).$
This in turn was motivated by the structure of the quark mass matrices. In
particular the near equality of the $(2,2)$ and $(2,3)$ Yukawa matrix
elements followed from the {\it smallness} of $V_{cb}.$ The possibility that
the $(2,2)$ and $(2,3)$ matrix elements should be equal inleading order has
been suggested by the recent data, particularly on the $B_{s}$ lifetime\cite
{Roberts:2001zy}.

\subsection{Soft Mass Terms}

One of the main motivations for a non-Abelian family symmetry is the need to
solve the flavour problem in supersymmetric models, suppressing large
contributions to FCNC coming from virtual diagrams involving non-degenerate
squark and slepton masses. Here we consider whether the $SU(3)$ family
symmetry is able to achieve this. Generically soft mass terms arise from
D-terms of the form $(\psi _{i}^{\dagger }\psi _{i}S^{\dagger }S)_{D}$ where 
$S$ is some supersymmetry breaking singlet field which has a non-vanishing
F-term $F_{S}$, leading to soft scalar masses $|F_{S}|^{2}(\tilde{\psi}%
_{i}^{\dagger }\tilde{\psi}_{i})/M^{\prime 2}$. The $SU(3)$ family symmetry
therefore ensures that the leading order soft scalar masses are proportional
to the unit matrix in family space. However the $SU(3)$ breaking Higgs
fields will lead to important corrections to the soft masses. The leading
correction is obtained from $((\psi _{i}\phi _{3}^{i})(\psi _{j}\phi
_{3}^{j})^{\dagger }S^{\dagger }S)_{D}/M^{\prime 4}.$ This leads to a
contribution to the third family soft scalar masses suppressed by a factor $%
O(a_{3}/M^{\prime })$. The second family receives soft scalar mass
corrections from $((\psi _{i}\phi _{23}^{i})(\psi _{j}\phi
_{23}^{j})^{\dagger }S^{\dagger }S)_{D}/M^{\prime 4}$ corresponding to a
suppression factor of $O(b/M^{\prime }).$ Here we have allowed for a new
mass scale, $M^{\prime },$ associated with these operators.

The detailed dynamical origin of the fermion mass operators discussed above
may play a role in determining the magnitude of $M^{\prime }.$ For example
if the fermion mass operators originate from a Higgs messenger sector, then
the higher order corrections to the leading soft operators could be highly
suppressed, $M^{\prime }>>M$ because they do not involve the Higgs sector.
If they originate from fermion messenger sectors then, since the expansion
parameter is larger in the down sector than in the up sector, this may lead
to right-handed scalar down mass corrections dominating over right-handed
scalar up mass corrections, and so on. Provided $M^{\prime }\geq M^{u},$ the
FCNC should be adequately suppressed due to the underlying $SU(3)$ symmetry.

The $SU(3)$ breaking effects can also give important corrections to vacuum
alignment. Recall that the equality of the soft mass components $%
m_{23,2}^{2}\phi _{23,2}^{2}=m_{23,3}^{2}\phi _{23,3}^{2}$ was crucial in
obtaining the vacuum $\phi _{23}=(0,b,b)$. The operator 
\begin{equation}
\frac{1}{M^{\prime \prime 4}}((\phi _{23,i}\overline{\phi _{3}^{i}})(\phi
_{23,j}\overline{\phi _{3}^{j}})^{\dagger }S^{\dagger }S)_{D}
\label{softphi}
\end{equation}
can give large corrections to the third component soft mass $m_{23,3}$
spoiling the vacuum alignment mechanism. Allowing for this splitting one
finds $\phi _{23}=(0,b_{2},b_{3})$ where $%
b_{2}/b_{3}=m_{23,2}^{2}/m_{23,3}^{2}$. In order for the vacuum alignment to
be preserved it is necessary for $a_{3}/M^{\prime \prime }$ to be small.
Whether this is the case depends on the details of the heavy messenger
sector.

In summary $SU(3)$ offers an elegant mechanism to make the soft quark and
lepton mass terms sufficiently close in mass to avoid unacceptably large
contributions to FCNC. Whether the SU(3) protection mechanism is
sufficiently robust depends on the details of the heavy messenger sector.

\section{Summary and Conclusions}

In this paper we have constructed a model of fermion masses based on an $%
SU(3)$ family symmetry. The novel feature of this implementation is the use
of the underlying $SU(3)$ to align the vacuum expectation values of the
fields spontaneously breaking $SU(3).$ This provides a mechanism for
explaining maximal mixing in the atmospheric neutrino sector in the case of
just three light neutrinos. Our model predicts either the LOW or the
quasi-vacuum solar solutions and a CHOOZ angle of order $V_{ub}.$ Allowing
only for unknown Yukawa couplings of $O(1),$ we were able simultaneously to
fit the quark and lepton masses and mixing angles in a model in which all
left-handed and charge conjugate right- handed fermion states belonged to $%
SU(3)$ triplets. Given that the lepton mixing angles are large while the
quark mixing angles are small it is remarkable it is possible to achieve
such a degree of similarity between quarks and leptons. The reason is due to
the structure of the neutrino mass matrix and the see-saw mechanism which
provides the mechanism to obtain large lepton mixing angles from the
structure responsible for small quark mixing angles. A bonus of the scheme
is the protection the $SU(3)$ family symmetry offers against large FCNC.

The model we have constructed is the low energy effective theory coming from
Beyond the Standard Model physics. While it is of interest to constuct the
underlying theory, it is clear that there is considerable flexibility in its
structure. For example we can construct a Higgs messenger model in which all
the higher dimension operators responsible for the light fermion masses
comes from mixing with heavy Higgs states. Alternatively there are examples
in which an underlying string theory gives rise to the same effective low
energy structure discussed here. We hope to address these questions in a
future publication.

\section*{Appendix - Vacuum Alignment}

Following Section \ref{vac} we start with the vacuum structure 
\begin{equation}
<\phi _{3}>=\left( 
\begin{array}{c}
0 \\ 
0 \\ 
a_{3}
\end{array}
\right) ,\ \ <\overline{\phi _{2}}>=\left( 
\begin{array}{c}
0 \\ 
a_{2} \\ 
0
\end{array}
\right) .
\end{equation}
which is to be triggered by radiative corrections. We assume that $m_{3}^{2}$
becomes negative at a much larger scale than $m_{\bar{2}}^{2}$ so that
ultimately we will have $a_{3}^{2}\gg a_{2}^{2}$. For radiativre corrections
to trigger a large symmetry breaking scale in a supersymmetric theory the
VEVs must develop along D-flat directions. Suppose there are scalar fields
transforming as $\overline{3}$ under $SU(3)$ and that $\overline{\phi _{3}}$
has the smallest soft mass amongst these. Then if $m_{{3}}^{2}(\Lambda
^{2})+m_{\bar{3}}^{2}(\Lambda ^{2})<0$ at some scale $\Lambda $ then a VEV
for $\phi _{3}$, $\overline{\phi _{3}}$ will appear and to leading
order D-flatness will align it with $<\phi _{3}>$, 
\begin{equation}
<\overline{\phi _{3}}>=\left( 
\begin{array}{c}
0 \\ 
0 \\ 
a_{3}
\end{array}
\right) .
\end{equation}
where $a_{3}\simeq \Lambda .$

Consider now the additional fields $\phi _{23}$, $\overline{\phi _{23}}$.
Throughout the analysis the important effect D-terms must be taken into
account, since these terms play an important role in determining which VEVs
occurs and in vacuum alignment. We assume that $\phi _{23}$ has a positive
mass squared and its VEV is triggered by minimising $F_{Y}$ from Eq.\ref{P2}
which leads to $<\phi _{23,2}\phi _{23,3}>=\mu ^{4}/(a_{2}a_{3})\equiv b^{2}$%
. The mass term $m_{23}^{2}\phi _{23}^{2}=m_{23}^{2}(\phi _{23,1}^{2}+\phi
_{23,2}^{2}+\phi _{23,3}^{2})$ is then minimised by equal VEVs in the 2 and
3 directions, 
\begin{equation}
<{\phi _{23}}>=\left( 
\begin{array}{c}
0 \\ 
b \\ 
b
\end{array}
\right) .
\end{equation}
Through the choice of the $\mu $ parameter in Eq.\ref{P2} we can arrange a
hierarchy $a_{3}\gg b\gg a_{2}$. Then $a_{2}$ can be made small enough so as
not to affect the mass matrices significantly.

Assuming $m_{\bar{3}}^{2}<m_{\bar{23}}^{2}<0,$ radiative symmetry breaking
triggers a VEV for $<\overline{\phi _{23}}>$ and minimising $F_{U}$ from Eq.%
\ref{P2} aligns it to be orthogonal to $<{\phi _{23}}>$, 
\begin{equation}
<\overline{\phi _{23}}>=\left( 
\begin{array}{r}
0 \\ 
b^{\prime } \\ 
-b^{\prime }
\end{array}
\right) .
\end{equation}

In order to determine $b^{\prime }$ we must allow for subleading corrections
to the D-flatness conditions which align $<\overline{\phi _{3}}>$ with $%
<\phi _{3}>\footnote{%
Note $x,y\ll a_{3}$ should not be confused with the fields $X,Y$ which we
have zero VEVs from F-flatness.}$, 
\begin{equation}
<\overline{\phi _{3}}>=\left( 
\begin{array}{c}
0 \\ 
y \\ 
a_{3}+x
\end{array}
\right) .
\end{equation}
Note that the $F_{Y}$ correction to $\phi _{23}$ VEV is proportional to $%
a_{2}$ and hence negligible.

The D-terms coming from the generators $T_{3}=diag(1,-1,0)$, $%
T_{8}=diag(1,0,-1)$, $T_{23}=\delta _{i2}\delta _{j3}$, are: 
\begin{eqnarray}
|D_{3}|^{2} &=&|-y^{2}+b^{2}-a_{2}^{2}-b^{\prime 2}|^{2}=0 \\
|D_{8}|^{2} &=&|a_{3}^{2}+b^{2}-(a_{3}+x)^{2}-b^{\prime 2}|^{2}=0 \\
|D_{23}|^{2} &=&|b^{2}+b^{\prime 2}-y(a_{3}+x)|^{2}=0
\end{eqnarray}
Solving for these we find, to leading order in $b/a_{3}$, that $x=0$, $%
y=2b^{2}/a_{3}$, $b^{\prime }=b$, which leads to the VEVs in Table \ref{vevs}%
.

By a similar analysis one can check that the preferred vacuum has zero VEVs
in the first components. Allowing for $<\overline{\phi _{3,1}}>=z$, $<{\phi
_{23,1}}>=\alpha $, $<\overline{\phi _{23,1}}>=\bar{\alpha}$, and
considering the same D-terms as previously and in addition those associated
with the generators $T_{12}=\delta _{i1}\delta _{j2}$, $T_{13}=\delta
_{i1}\delta _{j3}$, and also considering the soft mass terms associated with
the scalars $\overline{\phi _{3}}$, ${\phi _{23}}$, $\overline{\phi _{23}}$,
it can be shown that for $m_{23}^{2}>2m_{\overline{23}}^{2}$ that $\overline{%
\alpha }=\alpha =z=0$, to leading order.

\end{document}